\documentclass[twocolumn,aps,prb,superscriptaddress,unsortedaddress,showpacs]{revtex4}
\usepackage{graphicx}
\usepackage{epsf}

\begin{document}

\title{Transport implications of Fermi arcs in the pseudogap phase of the cuprates}

\author{A. Levchenko}
\affiliation{Materials Science Division, Argonne National Laboratory, Argonne, IL 60439}
\author{T. Micklitz}
\affiliation{Dahlem Center for Complex Quantum Systems and Institut f\"ur Theoretische Physik,
Freie Universit\"at Berlin, 14195 Berlin, Germany}
\author{M. R. Norman}
\affiliation{Materials Science Division, Argonne National Laboratory, Argonne, IL 60439}
\author{I. Paul}
\affiliation{Institut N\'eel, CNRS/UJF, 25 avenue des Martyrs, BP 166, 38042 Grenoble, France}

\begin{abstract}
We derive the fermionic contribution to the longitudinal and Hall
conductivities within a Kubo formalism, using a phenomenological
Greens function which has been previously developed to describe
photoemission data in the pseudogap phase of the cuprates. We find
that the in-plane electrical and thermal conductivities are
metallic-like, showing a universal limit behavior characteristic of
a d-wave spectrum as the scattering rate goes to zero.  In contrast,
the c-axis resistivity and the Hall number are insulating-like, being
divergent in the same limit.  The relation of these results to
transport data in the pseudogap phase is discussed.
\end{abstract}
\date{\today}
\maketitle

Photoemission data for underdoped cuprates reveal that the Fermi
surface in the pseudogap phase breaks up into disconnected Fermi
segments, known as Fermi arcs.\cite{Nat98} The length of these arcs
scales linearly with $T/T^*(x)$, where $T^*(x)$ is the doping
dependent pseudogap temperature.\cite{Amit}  This behavior can be
reproduced if one assumes a temperature independent d-wave gap with
a scattering rate proportional to $T$.\cite{Tallon,Norm07}

Obviously, such temperature dependent arcs should have profound
implications for the nature of the transport in the pseudogap
phase.\cite{Ando}  To investigate this, we take a Greens function
based on a phenomenological self-energy that generates these
arcs,\cite{PRB98,Norm07} and then construct the Kubo bubble.  We use
this to calculate both the in-plane and c-axis longitudinal
conductivities, as well as the Hall conductivity.  We then connect
our results to transport data for the cuprates.

The in-plane conductivity at T=0 is given by the Kubo formula \cite{Mahan}
\begin{equation}
\sigma_{xx} = \frac{2 e^2\hbar}{\pi} \int \frac{d^3k}{(2\pi)^3}
v_x^2(k) [\mathrm{Im}G(k,0)]^2
\end{equation}
where $v_x$ is the $x$ component of the Fermi velocity,
$\mathrm{Im}G(k,0)\equiv\mathrm{Im}G^R(k,\omega=0)$ with $G^R$ the retarded
Greens function, and the factor of 2 comes from summing over spin.
Transforming the $k$ integral to one over $\epsilon$ and $\phi$, and ignoring the
$\phi$ dependence of $v$ and any c-axis dispersion, we have the planar conductivity
\begin{equation}
\sigma = \frac{e^2\hbar N v^2}{\pi d} \int d\epsilon
\frac{d\phi}{2\pi} [\mathrm{Im} G(\epsilon,\phi,0)]^2
\end{equation}
where $N$ is the (two dimensional) normal state density of states (per spin), $v$ the Fermi velocity,
and $d$ the c-axis thickness divided by the number of conducting planes.

We now consider a BCS model for the Greens function.
In the so-called `single lifetime' version,\cite{PRB98,Tallon,Norm07}
\begin{equation}
-\mathrm{Im}G(\epsilon,\phi,0) =
\frac{\Gamma}{\epsilon^2+\Gamma^2+\Delta^2(\phi)}
\end{equation}
where $\Delta$ is the pairing gap and $\Gamma$ the inverse lifetime.\cite{foot1}
This form for $G$ gives a good description of photoemission data in the pseudogap phase.
In particular, if $\Gamma$ scales as $T$, then the $T$ dependence of the arc length is
reproduced, as well as the variation of the spectral gap around the Fermi surface.

Substituting this form of $G$ into the expression for $\sigma$, the $\epsilon$ integral
is convergent, and setting its limits to infinity yields
\begin{equation}
\sigma  =  \frac{e^2\hbar N v^2 \Gamma^2}{\pi d} \int_0^{\pi/2} \frac{dx}{(\Gamma^2+\Delta_0^2\cos^2x)^{3/2}}
\end{equation}
where in the d-wave case, $\Delta(\phi)=\Delta_0\cos x$ with $x=2\phi$.
Performing the $x$ integration
\begin{equation}
\sigma  = \frac{e^2\hbar N v^2\lambda}{\pi d\Delta_0}E(\lambda)
\end{equation}
where $\lambda = \Delta_0/\sqrt{\Gamma^2+\Delta_0^2}$ and
$E$ is the complete elliptic integral of the second kind.

This can be easily generalized \cite{Ioffe} to the case of the c-axis conductivity by replacing $v^2/2$
by $t^2_{\perp}(\phi)d^2/\hbar^2$, where $t_{\perp}$ is the interlayer tunneling energy whose angle dependence goes
as $\cos^2(2\phi)$.\cite{PWA,Ole}  This has a profound effect on the conductivity,\cite{LeeRMP}
\begin{equation}
\sigma_c =  \frac{2e^2d N t_{\perp}^2 \Gamma^2}{\pi\hbar}
\int_0^{\pi/2}
\frac{dx\cos^4(x)}{(\Gamma^2+\Delta_0^2\cos^2x)^{3/2}}
\end{equation}
Evaluating, one finds
\begin{equation}
\sigma_c = \frac{2e^2d N t_{\perp}^2 \Gamma^2}{\pi\hbar\Delta_0^3\lambda}
[(2-\lambda^2)E(\lambda) - 2(1-\lambda^2)K(\lambda)]
\end{equation}
where $K$ is the complete elliptic integral of the first kind.

In Fig.~1, the inverse of $\sigma$ ($\rho$) and $\sigma_c$
($\rho_c$) are plotted as a function of $\Gamma$. For zero
$\Delta_0$, they are proportional to $\Gamma$ as expected.  For
non-zero $\Delta_0$, it is seen that $\rho$ saturates to a finite
value in the limit that $\Gamma$ goes to zero.  This is the
universal limit discussed by Patrick Lee.\cite{Lee}  The
resistance has a weak minimum when $\Gamma/\Delta_0 \sim 0.46$,
before increasing like $\Gamma$ as in the normal state. The c-axis
conductivity behaves quite differently because of the
$\cos^4(2\phi)$ factor, which is peaked at $\phi=0$ where the d-wave
gap is maximal (thus killing off the universal behavior associated
with the nodes).  As a consequence, $\rho_c$ diverges as $1/\Gamma^2$
(this behavior coming from the prefactor in Eq.~6). As $\Gamma$
increases, a strong minimum is seen in $\rho_c$ at $\Gamma/\Delta_0
\sim 1.26$, before increasing like $\Gamma$.

\begin{figure}
\centerline{\includegraphics[width=3.4in]{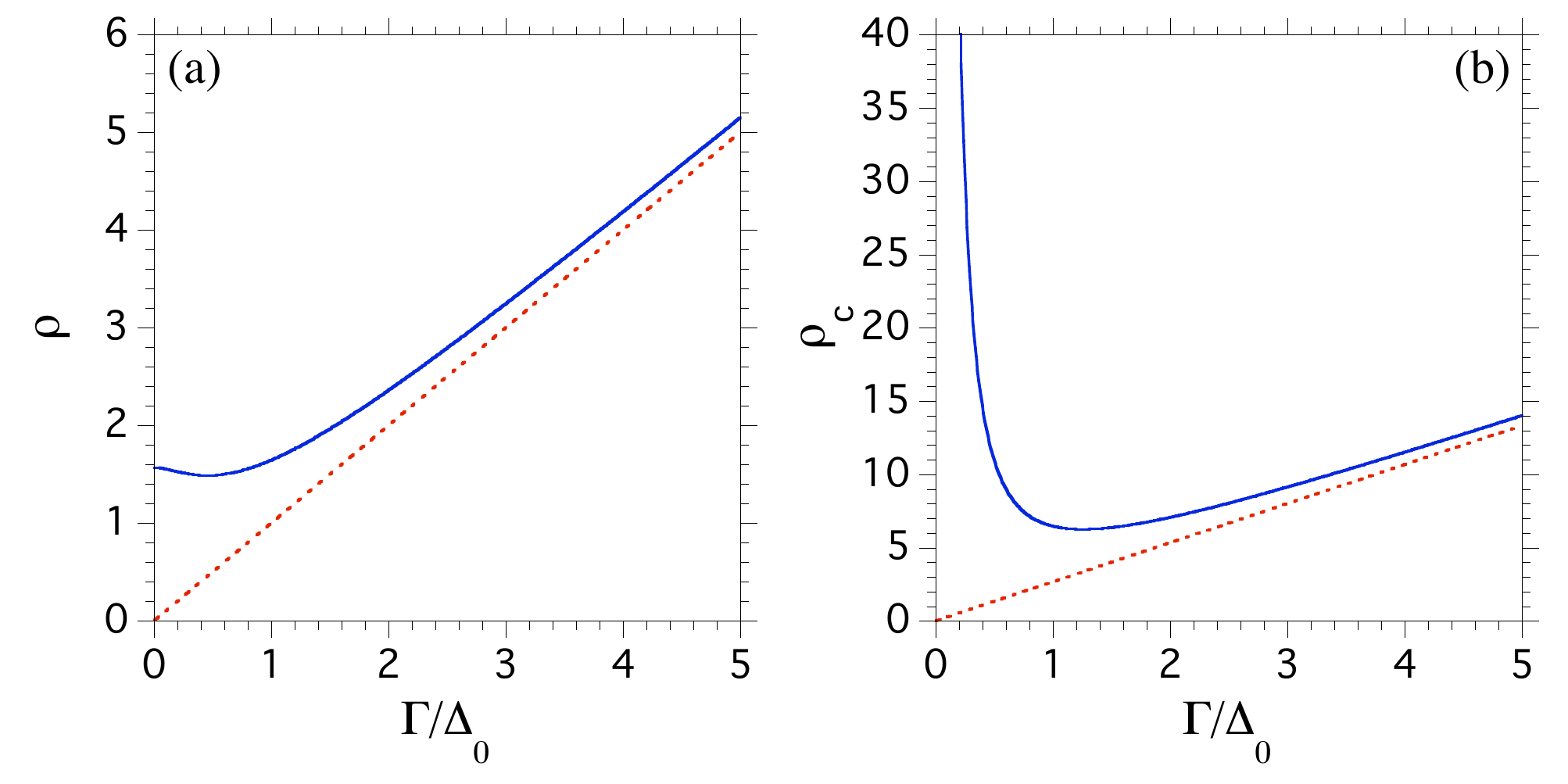}} \caption{(Color
online) (a) planar and (b) c-axis resistivity versus $\Gamma$.  The
units for $\rho$ are $2m^*/(e^2\hbar n)$ and for $\rho_c$ are
$\hbar^2v^2m^*/(d^2e^2\hbar n t_{\perp}^2)$, where $n/m^*=Nv^2/d$ with $n$ the
electron density and $m^*$ the effective mass. The dashed lines are
for $\Delta_0=0$.} \label{fig1}
\end{figure}

Although the above results are for $T$=0, we have done numerical studies which included the
thermal factors in the Kubo bubble.  Only minor differences were seen, and thus to a good
approximation, the above $T$=0 formulas are sufficient.  Therefore, if $\Delta_0$ is constant,
and $\Gamma$ scales as $T$, as commonly assumed to describe the photoemission data, then
the x-axis of the above figures can be read as temperature.

We now turn to the Hall conductivity (current in the plane, field along the c-axis) which is easily
derived by insertion of a magnetic field vertex into the Kubo bubble \cite{foot2}
\begin{equation}
\sigma_H =-\frac{2e^2\hbar^2 N v^2 \omega_c}{3\pi d} \int d\epsilon
\frac{d\phi}{2\pi} [\mathrm{Im} G(\epsilon,\phi,0)]^3
\end{equation}
where $\omega_c$ is the cyclotron frequency ($eH/m^*c$).
Substituting $G$ from Eq.~3 and performing the integral over $\epsilon$
\begin{equation}
\sigma_H =  \frac{e^2\hbar^2 N v^2 \Gamma^3 \omega_c}{2\pi d}
\int_0^{\pi/2} \frac{dx}{(\Gamma^2+\Delta_0^2\cos^2x)^{5/2}}
\end{equation}
Evaluating, one finds
\begin{equation}
\sigma_H  = \frac{e^2\hbar^2 N v^2 \omega_c\lambda}{6\pi d\Gamma\Delta_0}
[2(2-\lambda^2)E(\lambda)-(1-\lambda^2)K(\lambda)]
\end{equation}
Note that the Hall resistivity ($\rho_H$)
is $\sigma_H/\sigma^2$, and the Hall coefficient ($R_H$) is $\rho_H/H$.

\begin{figure}
\centerline{\includegraphics[width=3.4in]{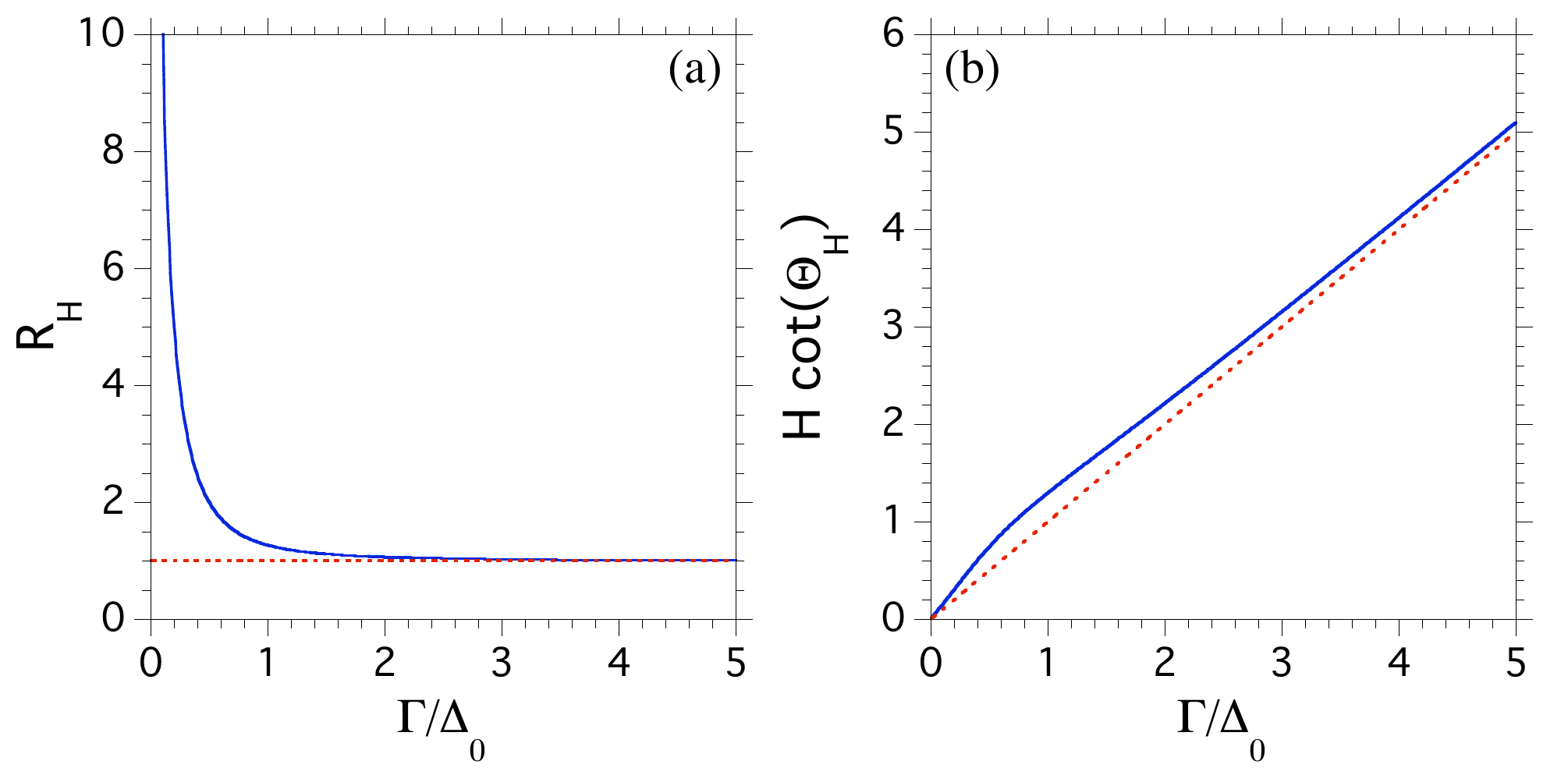}} \caption{(Color
online) (a) Hall coefficient and (b) cotangent of the Hall angle
versus $\Gamma$. The units for $R_H$ are $1/(nec)$ and
$H\cot(\Theta_H)$ are $2cm^*/(e\hbar)$. The dashed lines are for
$\Delta_0=0$.} \label{fig2}
\end{figure}

In Fig.~2a, we plot the Hall number as a function of $\Gamma$, where
it is seen to diverge as $1/\Gamma$ at small $\Gamma$.  This
behavior is due to the nodal contribution, thus generalizing the
results of Ref.~\onlinecite{Lee} to the Hall conductivity. In
Fig.~2b, we plot the cotangent of the Hall angle,
$H\cot(\Theta_H)=\rho/R_H$, which vanishes as $\Gamma$ goes to zero,
but increases as $\Gamma$ for large $\Gamma$.

These results are easily generalized to their thermal counterparts.\cite{Paul}  The in-plane thermal
conductivity is
\begin{equation}
\kappa_{xx} = \frac{2 \pi \hbar k_B^2 T }{3} \int
\frac{d^3k}{(2\pi)^3} v_x^2(k) [\mathrm{Im}G(k,0)]^2
\end{equation}
and thus we recover \cite{Durst} the Wiedemann-Franz law $\kappa =
\frac{\pi^2k_B^2T\sigma}{3e^2}$ (therefore, $\kappa$ exhibits
universal behavior as well). For the thermopower, $S$, and the
Nernst, one must consider particle-hole asymmetry effects.  In our
simple model where $\Gamma$ is taken as a momentum and frequency
independent constant, the only source for this is the density of
states, $N$. As a consequence, we find that $S/T$ is a constant
($\frac{\pi^2 k_B^2}{3e}\frac{d\ln N}{d\epsilon}$), and that the
Nernst effect vanishes due to the Sondheimer cancellation (i.e.,
$\frac{\partial \tan(\Theta_H)}{\partial \epsilon} = 0$).  These
results will obviously change if a more sophisticated model is used
for the self-energy.\cite{Paul}

We now discuss implications of our results.  The first point we wish to make is almost trivial.
That is, transport data for various dopings scale as a function of $T/T^*(x)$.\cite{Wuyts,Luo}
Since $\Delta_0$ scales with $T^*(x)$ \cite{JC99} and $\Gamma$ with $T$,\cite{MFL} then our
transport results also scale as $T/T^*(x)$.  Within our model, the scaling factor
is such that $\Gamma = \sqrt{3} \Delta_0 T/T^*$, noting that
$\Gamma/\Delta_0 > \sqrt{3}$ is the condition for gaplessness of
$\mathrm{Im}G(\epsilon=0,\phi=0,\omega)$.\cite{Norm07}
Since $\Delta_0/T^* \sim 2$, this reduces to $\Gamma \sim 2 \sqrt{3} T$.

Now, as the arc length scales with $\Gamma$, one might have naively
expected that the planar resistivity would diverge as $\Gamma$ goes
to zero. This does not occur for the same reasons discussed by
Lee.\cite{Lee}  That is, for a d-wave spectral gap, the residual
conductivity is independent of the scattering rate.  This result,
though, does not hold for the c-axis resistivity, which indeed
diverges.  Experimentally, the in-plane resistivity is indeed
metallic-like in the pseudogap phase, whereas the c-axis resistivity
is divergent.\cite{Timusk}  So, this basic dichotomy of the cuprates
is trivially explained by our model. On the other hand, the
experimental in-plane resistance below $T^*$ falls below that of the
high temperature linear $T$ behavior of the normal state, whereas
our model results fall above.  This indicates that an extra bosonic
contribution to the conductivity should exist, a likely source being
the pairs themselves.  In fact, it is well known that there is a
significant contribution to the conductivity above $T_c$ which
follows the 2D Aslamazov-Larkin form.\cite{Leridon}  We will
investigate these bosonic contributions in a future
paper.\cite{Alex} In regards to the c-axis resistance, the data are
usually fit by an activated form,\cite{Luo} rather than the power
law we find.  Our results, though, are obviously dependent on the
precise form of $t_{\perp}(\phi)$ and $\Delta(\phi)$, and also to any temperature
dependence of $\Delta_0$ and $t_{\perp}$.  Inclusion of impurity scattering
will also cut off the divergence.

We now turn to the Hall conductivity.  Our results are roughly
consistent with the reported variation of $R_H$ versus temperature
in the pseudogap phase,\cite{Hwang,Zoritsa,Ando2} though our expression
is more singular than the data.  The simple function $\sqrt{a^2 +
b^2/\Gamma^2}$ does a good job of fitting the curve in Fig.~2a.
The above caveats about the temperature dependence of $\Delta_0$,
the inclusion of impurity scattering, and bosonic contributions to
the conductivity should be kept in mind.  In regards to Fig.~2b, the
actual Hall angle scales as $T^2$ rather than $T$ as we find,
indicating different lifetimes entering $\sigma_H$ and $\sigma$ as
has been previously commented on.\cite{Chien} In that context, we
note that $\Gamma$ in principle can be a function of angle, and that
its temperature variation, as well as that of $\Delta_0$, can also
be angular dependent.  Some evidence for this has been provided by
photoemission,\cite{Ming} tunneling,\cite{Alldredge} and transport
studies.\cite{Hussey}  Finally, we note that a related study to ours
was recently done by Smith and McKenzie,\cite{Smith} where they
considered other model Greens functions discussed in
Ref.~\onlinecite{Norm07} as well.

In conclusion, we have calculated the temperature variation of
various transport quantities within a simple model previously used
to describe photoemission data in the pseudogap phase.  We find that
the in-plane electrical and thermal conductivities are
metallic-like, but the c-axis and Hall conductivities are
insulating-like, in qualitative agreement with the experimental
data.

This work was supported by the U.S. DOE, Office of Science, under contract DE-AC02-06CH11357.


\begin{thebibliography}{99}
\bibitem{foot1}
As we are above $T_c$, then by $\Delta^2$ we mean $<\Delta^2>$.
\bibitem{foot2}
The sign reflects that the Fermi surface is hole-like.
\bibitem{Durst}
We do not include an anomalous correction to the thermal current, as was done in
A. C. Durst and P. A. Lee, Phys. Rev. B {\bf 62}, 1270 (2000), since we assume
that $<\Delta>$=0 because we are above $T_c$.  As a result, we obey
the Wiedemann-Franz law.
\end{thebibliography}
\end{document}